\documentclass[aps,superscriptaddress,twocolumn,pra]{revtex4-1}
\usepackage{graphicx}
\usepackage{amssymb}
\usepackage{amsmath}
\usepackage{bm}
\usepackage{enumerate}
\usepackage{verbatim}
\usepackage[bookmarks=false,colorlinks=true,urlcolor=blue,citecolor=blue,linkcolor=blue]{hyperref}
\usepackage[normalem]{ulem}
\usepackage{epstopdf}
\def\ep{\varepsilon}

\begin{document}

\title{Localization Lifetime of a Many-Body System with Periodic Constructed Disorder}

\author{Michael Schecter}
\affiliation{Center for Quantum Devices, Niels Bohr Institute, University of Copenhagen, 2100 Copenhagen, Denmark}

\author{Michael Shapiro}
\affiliation{Department of Mathematics, Michigan State University, East Lansing, Michigan 48824, USA}

\author{Mark I. Dykman}
\affiliation{Department of Physics and Astronomy, Michigan State University, East Lansing, Michigan 48824, USA}

\date{\today}

\begin{abstract}
We show that, in a many-body system, all particles can be strongly confined to the initially occupied sites for a time that scales as a high power of the ratio of the bandwidth of site energies to the hopping amplitude. Such time-domain formulation is complementary to the formulation of the many-body localization of all stationary states with a large localization length. The long localization lifetime is achieved by constructing a periodic sequence of site energies with a large period in a one-dimensional chain. The scaling of the localization lifetime is independent of the number of particles for a broad range of the coupling strength. The analytical results are confirmed by numerical calculations.

\end{abstract}

\maketitle

\section{Introduction}\label{sec:intro}

Recently, a growing body of evidence for the existence of a many-body localized (MBL) phase has emerged \cite{Basko2006,Oganesyan2007,Pal2010a,Huse2013,Altshuler2016}, and now it includes a rigorous proof for a physically reasonable one-dimensional lattice model \cite{Imbrie2016} as well as experimental signatures from cold atoms in optical lattices \cite{Schreiber2015,Choi2016} and trapped ions \cite{Smith2016,Zhang2016}. A distinguishing feature of the MBL phase is a complete set of local integrals of motion \cite{Serbyn2013,Huse2014}.  The existence of such integrals implies that the system retains memory of the initial conditions and external manipulations imposed on it. This naturally leads to 
a potential use of the MBL phase for quantum information processing \cite{Choi2015}, and in particular for storing quantum information \cite{Smith2016}.

On the practical side, a drawback for applications of the MBL phase is that local integrals, sometimes referred to as ``l-bits", may have small overlap with the physically observable/manipulatable degrees of freedom (``p-bits"), e.g. the on-site occupation numbers or spin states. From the point of view of quantum control or quantum computing, one desires a strong overlap between l-bits and p-bits to ensure high fidelity of physically addressable quantities. In particular, in a quantum computer one requires all excitations initialized on-site to remain there for a sufficient time, typically given by a coherence time set by residual coupling to the environment or noise. The site localization, which we shall colloquially refer to as {\em confinement}, is a stronger constraint than the exponential decay of the wavefunction at large distances. 

Confinement of all states does not occur in a system with a random or incommensurate potential even for nearest neighbor hopping that we consider in this paper. This is because the probability to find two neighboring sites with energy difference smaller than the hopping amplitude $J$ is $\sim J/h$, where $h\gg J$ is the bandwidth of site energies and the distribution of the site energies is assumed uniform. As a result the excitations will hybridize between the sites over a time $\propto 1/J$  with probability $J/h$ per excitation \cite{Georgeot2000}.

\begin{figure*}[t!]
\centering
\includegraphics[width=2.0\columnwidth]{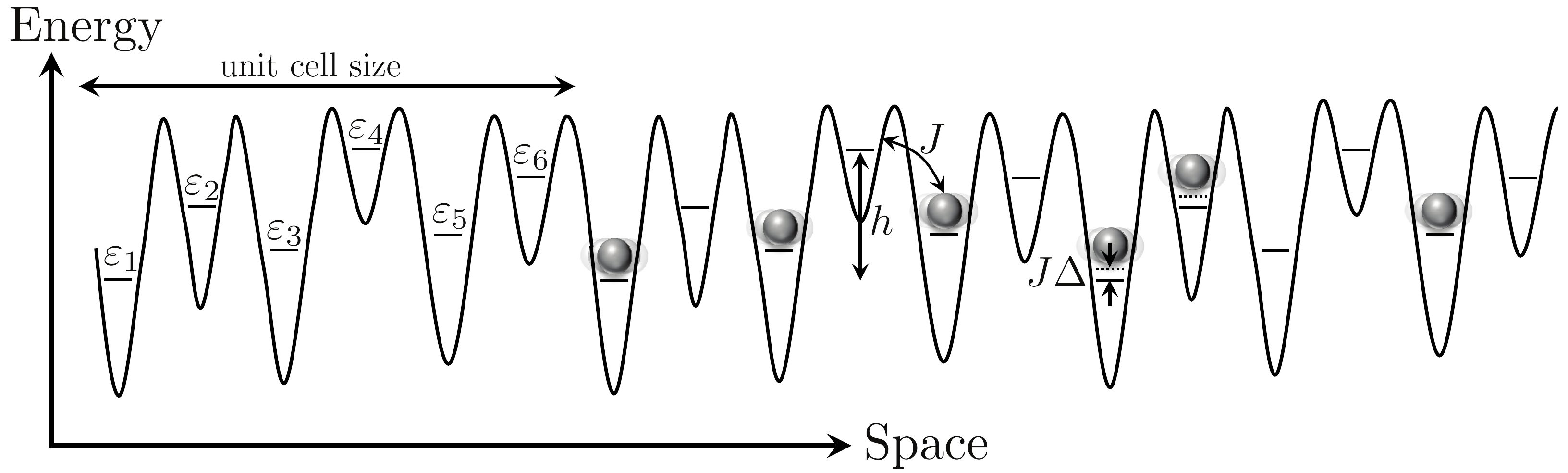}
\caption{Periodic potential whose unit cell is designed to enhance the on-site localization lifetime of a strongly interacting many-body system through the suppression of many-particle (many-excitation)  resonances. The wave functions of particles (gray spheres), which are  initially localized on a set of sites with site energies $\ep_i$, weakly spread onto neighboring sites, but  remain strongly confined for a time $t^*\propto J^{-1} (h/J)^{\alpha-1}$, cf. Eq.~\eqref{eq:bound}, where $\alpha$ is the number of sites per period; in the considered case $\alpha=6$. The bandwidth of the site energies is $h=\max_{i,j} |\ep_i-\ep_j|$. The site energy is shifted by $J\Delta$ if a neighboring site is occupied because of the particle-particle interaction.}
\label{fig:schematic}
\vspace{-0.0 in}
\end{figure*}

For non-interacting excitations, resonant hopping can be suppressed by explicitly {\em constructing} an appropriate ``disorder" of the site energy sequence \cite{Santos2005}. This is done for a finite bandwidth of the site energies by successively tuning them so that the further away the sites are the smaller is the detuning. One can thus achieve near unity l-bit/p-bit overlap of an infinite number of states.

The situation is significantly complicated by the presence of many-body interactions even where the interaction is short-range. This is a consequence of the exponential increase of the number of states with the number of excitations. One could therefore expect it to become impossible to fully suppress all many-body resonances in a large system for a finite site energy bandwidth, that is, to have the energy difference between different distributions of the excitations over sites larger than the matrix element of hopping between these distributions. We emphasize that this condition is not required in the MBL. The strong many-body confinement of excitations on sites is a strictly transient phenomenon even in the presence of the MBL. 

In this paper we consider the problem of optimizing the time on which, in a many-body system, excitations remain closely bound to the sites on which they were prepared. We call this time the localization lifetime \cite{Santos2005}. The optimization is achieved by tuning site energies. Such tuning was recently implemented  in trapped ion systems in order to achieve the MBL \cite{Smith2016,Zhang2016,Lee2016}. However, the MBL is not required to optimize the localization lifetime.  The lifetime formulation is meaningful independent of the structure of stationary eigenstates. Our analysis is also different from the problem of time evolution in the MBL systems \cite{Khemani2015}. Further motivation for studying the localization lifetime comes from the fact that, in realistic systems, the bath coupling and noise cannot be fully turned off, and the eigenstate character of the closed system is relevant only for times smaller than the decay or decoherence times induced by the environment.

In terms of quantum information processing, it is important to maximize the ratio of the hopping amplitude to the bandwidth $J/h$  while maintaining a long localization lifetime. Parameter $J$ characterizes the rate of two-qubit operations, whereas $h$ determines the characteristic bandwidth of the control field (for example, radiation) coupled to qubits, which is usually limited in the experiment \cite{Reed2010}.  

Here we propose a construction of the disorder that enables having the localization lifetime that scales as a high power of the large parameter $(J/h)^{-1}$. A key observation that makes our approach different from that in Ref.~\cite{Santos2005} is that this can be accomplished by using a one-dimensional chain with a {\bf periodic} sequence of site energies. The periodicity grossly simplifies the construction. It also brings in the translational symmetry, which allows one to think of the band structure of many-body excitations. 

The reciprocal widths of the many-excitation bands in a periodic system limits the many-excitation localization lifetime. In a sense, this is a weak constraint on the number of excitations involved, as the width quickly falls off as this number increases. However, the problem is not only to have the bands narrow, but also to eliminate many-excitation resonances within the period of the sequence. As we show, we can construct a periodic sequence of site energies in such a way that the scaling  of the localization lifetime with $J/h$ holds uniformly for any number of excitations for the period equal to 6 sites. The robustness of the sequence is guaranteed by the fact that the site energies are given by simple fractions of their bandwidth $h$ with a small common denominator. For the same reason, a strict periodicity of the sequence is not required either.

We consider a one-dimensional model of interacting spinless fermions, or equivalently an anisotropic Heisenberg spin one-half chain. It is described by the Hamiltonian
\begin{align}
\label{eq:H}
H=H_0+V, &\qquad H_0= \sum_j \varepsilon_j \hat{n}_j+J\Delta\sum_j \hat{n}_j \hat{n}_{j+1},\nonumber\\
&V= \frac{1}{2}J\sum_j c^\dagger_j c_{j+1}+{\rm h.c.},
\end{align}
where $c^\dagger_j$ and $c_j$ are fermionic creation and annihilation operators at site $j$, $\hat{n}_j=c^\dagger_j c_j$., and we set $\hbar =1$.  The site energies $\ep_j$ lie in the band $0\leq\varepsilon_j\leq h$ and form a period-six sequence, $\ep_{j+6} = \ep_j$, see Fig.~\ref{fig:schematic}. The formulation based on Eq.~(\ref{eq:H}) allows us to talk about particles rather than excitations, and this is the language we will be using in what follows.

As we show, by choosing $\ep_j$ we can achieve that, if we initially at $t=0$ prepare a configuration of particles in eigenstates of $\hat n_j$ such that $n_j(0) = 1$ for some sites $j$, then the population of these sites $n_j(t)\equiv \langle \hat{n}_j(t)\rangle_{n_j(0)=1}$ remains close to 1 for the localization lifetime $t^*$, 
\begin{eqnarray}
\label{eq:bound}
1- n_j(t) &<&\frac{10J^2}{h^2},\quad t\lesssim t^* = \frac{h^5}{50J^6}.
\end{eqnarray}
This means that, for example, for the ratio of bandwidth to hopping  $h/J=20$ we have $1- n_j(t) \lesssim0.025$ for $t\lesssim t^*\sim 10^5J^{-1}$. The parameter $t^*J$ can be compared with the ratio of the rate of a two-qubit operation to the decoherence rate, which is often smaller then $10^5$  in the current implementations of quantum computers. The estimates (\ref{eq:bound}) are applicable up to the regime of strong interaction $\Delta\lesssim 1$, where the interaction energy becomes comparable to the hopping amplitude $J$.

The remainder of the paper is organized as follows. In Sec.~\ref{sec:seq} we present our proposed energy sequence and explain its construction in the context of many-particle hopping. In Sec.~\ref{sec:single-particle} we consider the localization dynamics of a single particle initialized on a given site, and derive the localization lifetime. In Sec.~\ref{sec:many-particle} we incorporate the effects of interactions to calculate the leading order multi-particle transition amplitudes that give rise to off-site transport. Our analytic results for the infinite chain are supplemented by numerical diagonalization of the half-filled 12 site chain. Conclusions are provided in Sec.~\ref{sec:conclusion}.

\section{Energy Sequence Construction}\label{sec:seq} 

An advantageous feature of a periodic site energy sequence is that it allows one to characterize dynamic properties of an infinite chain by analyzing small segments on the order of the period. A basic problem is the account of many-particle resonances where different configurations $\{n_j\}$ have the same energy $\sum \ep_jn_j$. It is necessary to eliminate resonances between all single-particle inter-site transitions within the period, all two-particle transitions, etc. Such transitions come from the hopping term $\propto J$ in the Hamiltonian (\ref{eq:H}). An important characteristic is the number of single-particle ``steps" between neighboring sites $N_s$  involved in a resonant transition. Out of $N_s$ steps,  $N_s-1$ are virtual nonresonant transitions. 

In calculating $N_s$ one should take into account that a many-particle transition would not occur unless the particles interact with each other. This only happens when they are on neighboring sites. Therefore the steps must include bringing the particles to neighboring sites. As a result, the total amplitude of the many-particle transition contains a factor $\propto (J/h)^{N_s-1}$ and factor $J\Delta/h$ raised to the power that shows how many interactions are involved.

The period of the sequence $N$ gives the maximal number of steps for a resonant single-particle transition; in this case $N_s = N$. We now consider a potentially resonant two-particle transition from sites $(i,j)$ to sites $(i',j')$, with $\ep_i+\ep_j = \ep_{i'}+\ep_{j'}$, with all sites within the period $N$. The number of steps for such a transition is bound by $2N-6$, because the particles have first to make virtual transitions from $(i,j)$ to some neighboring sites $(p,p+1)$, where they interact, and then move away to $(i',j')$. A more detailed explanation and the analysis of the contribution from the transitions between sites separated by more than $N$ is given in Sec.~\ref{sec:many-particle}.

Our goal is to eliminate all many-particle resonances. As a first step, we want to do so within a period. The above estimate shows that, to eliminate resonances for all single- and two-particle transitions on equal footing, we should set $N=6$.  As we show later, for $N=6$ we can also eliminate 5-step resonances for a larger number of particles. As it turns out, here a critical consideration of 3-particle resonances must be made. We show that such resonances give a subleading contribution compared to single-particle resonances for $\Delta\lesssim 1$. These arguments make the period $N=6$ special.

To achieve robustness with respect to errors in the site energies, we choose these energies as {\it simple fractions} of the bandwidth $h$, $\ep_j=he_j/r$ with integer $e_j$ and $r$ and with $e_j\leq r$; without loss of generality we set $e_1=0$.  The magnitude of all distinct many-particle energy differences are bounded from below by $h/r$, where $r$ is the least common denominator of the energy sequence. The problem is then reduced to finding a proper set $\{e_j\}$, with minimal $r$, where all many-body energy differences are distinct. This is a formidable problem when many particles are considered, and $r$ is expected to grow with the increasing number of particles, leading to a loss in robustness. It also rapidly grows with the increasing period $N$, which provides another motivation for keeping $N$ not too large. 

In order to maximize the localization lifetime, for our period-6 sequence we choose the values of the site energies so that
\begin{enumerate}[(i)]
\item There are no one-particle, two-particle or three-particle resonances up to order $J^6$.
\item The difference between energies of neighboring sites should be maximized, and in each case exceed roughly one half the total bandwidth.
\end{enumerate}
These heuristic conditions together can provide a long localization lifetime $t^*$, see Eq.~(\ref{eq:bound}). Resonant transitions of an initially localized on-site state require multiple nonresonant steps, so that their amplitude starts from $J^6$. This appears to be true also for four and five-particle resonances and is trivially satisfied for resonances that involve moving more than 5 particles. 
  
Condition (i) requires that all numerators within a period must be distinct, $e_i\ne e_j$ for all $i\ne j,\ i,j\in[1,6]$ to avoid single-particle resonances. Similarly, the absence of two-particle resonances is equivalent to $e_i+e_j\ne e_k+e_l$ for all pairwise distinct $e_i,\,e_j,\,e_k,\,e_l$. Given that $e_1=0$, the single-particle constraint makes $r = (e_j)_{\max}$  to be at least $5$, but the many-particle constraints are much harder to satisfy. With these constraints, $r$ increases with $N$ superlinearly.   

A sufficient procedure to satisfy the condition of the absence of one- and two-particle resonances entails building the sequence $\ep_j$ term by term. For two sites the solution for the integer set $\{e_j\}$ with minimal $r$, written in ascending order of the elements, is \{0,\,1\} (with $r=1$), while for three and four sites it is $\{0,1,2\}$ (with $r=2$) and $\{0,1,2,4\}$ (with $r=4$), respectively. One may notice that the next term in the brackets of the sequence  can be obtained recursively in terms of the two prior terms in the brackets as $e_j=e_{j-1}+e_{j-2}+1$ for $j\geq 2$ with $e_0\equiv 0$. Such a sequence is a shifted version of the celebrated Fibonacci sequence $\mathcal{F}_j=\{1,\,1,\,2,\,3,\,5,\,8,\,13,\,\dots\}$, expressible as $e_j=\mathcal{F}_{j+1}-1=\{0,\,1,\,2,\,4,\,7,\,12\}$ for $j_{\max}\equiv N =6$. Respectively, $r=(e_j)_{\max}=12$ for $N=6$.

Although the structure of the shifted Fibonacci sequence is sufficient to guarantee the absence of all one-particle and two-particle resonances, it suffers from the three-particle resonance $e_1+e_2+e_6=e_3+e_4+e_5$ and would not satisfy (ii) upon any reordering. Remarkably, by modifying $e_{4}\to e_{4}+2$ and $e_{5}\to e_{5}+2$ we also eliminate all three-particle resonances in the unit cell, and thus (i) is satisfied. The sequence can be reordered to satisfy (ii). This gives 
\begin{equation}
\label{eq:seq}
\varepsilon_j =\frac{h}{12}\{0,\,6,\,1,\,12,\,2,\,9\}, \quad \varepsilon_j=\varepsilon_{j+6}
\end{equation}
 for $j=1,\ldots,6$. 

The above analysis assumes that no many-particle {\em inter-}peiod resonances arise, that is, all transitions happen between sites separated by no more than $N=6$ positions. While this is guaranteed by construction at the one-particle level, in Sec.~\ref{sec:many-particle} we verify that it is also true at the many-particle level as long as we are looking at the transition amplitude that scales with $J$ as $J^6$. 

The Hamiltonian (\ref{eq:H}) does not have particle-hole symmetry. Our numerical data refer to $\Delta >0$. The results for the localization lifetime in the case $\Delta <0$ coincide with the results for $\Delta>0$ with the same $|\Delta|$ if one uses instead of Eq.~(\ref{eq:seq})  the sequence $\{\varepsilon_j\}=\frac{h}{12}\{12,\,6,\,11,\,0,\,10,\,3\}$. It is obtained from Eq.~\eqref{eq:seq} under sign-change of each term and a constant energy shift.

\section{Single particle localization}\label{sec:single-particle}

We start by considering the localization lifetime for noninteracting particles. The single-particle dynamics is readily understood in terms of the stationary states of Eq.~\eqref{eq:H}.  Due to the periodicity of  Eq.~\eqref{eq:seq}, these states are Bloch states with dimensionless momentum $k$ and a band index. For the period $N=6$ this index takes on 6 values and $-\pi/6 \leq k\leq \pi/6$.  The band structure is most simple for large spacing between the site energies $\ep_j$, where the differences between the values of $\ep_j$ within a period $1\leq j \leq 6$  exceed the bandwidths $2|U_j|$. The Bloch bands can then be described in a tight-binding approximation, with individual bands centered at the site energies $\ep_j$,
\begin{equation}\label{eq:spec}
E_{j}(k)=\tilde{\varepsilon}_{j}+U_j\cos 6k,
\end{equation}
To the leading order in $J$ the renormalized site energies and the bandwidths have the form

\begin{align}
\label{eq:energy_shift}
&\tilde{\varepsilon}_{j}\simeq\varepsilon_{j}+\left(\frac{J}{2}\right)^{2}\sum_{\alpha=\pm 1}\frac{1}{\varepsilon_{j}-\varepsilon_{j+\alpha}}, \nonumber\\
&U_j= 2\left(\frac{J}{2}\right)^{6}\prod_{m=1}^{5}\frac{1}{\varepsilon_{j}-\varepsilon_{j+m}}\,.
\end{align}
For the site energy sequence \eqref{eq:seq} the largest bandwidth occurs for $j=3$, $|U_3|\approx 18 J\left(J/h\right)^{5}$.

\begin{figure}[t!]
\centering
\includegraphics[width=1.\columnwidth]{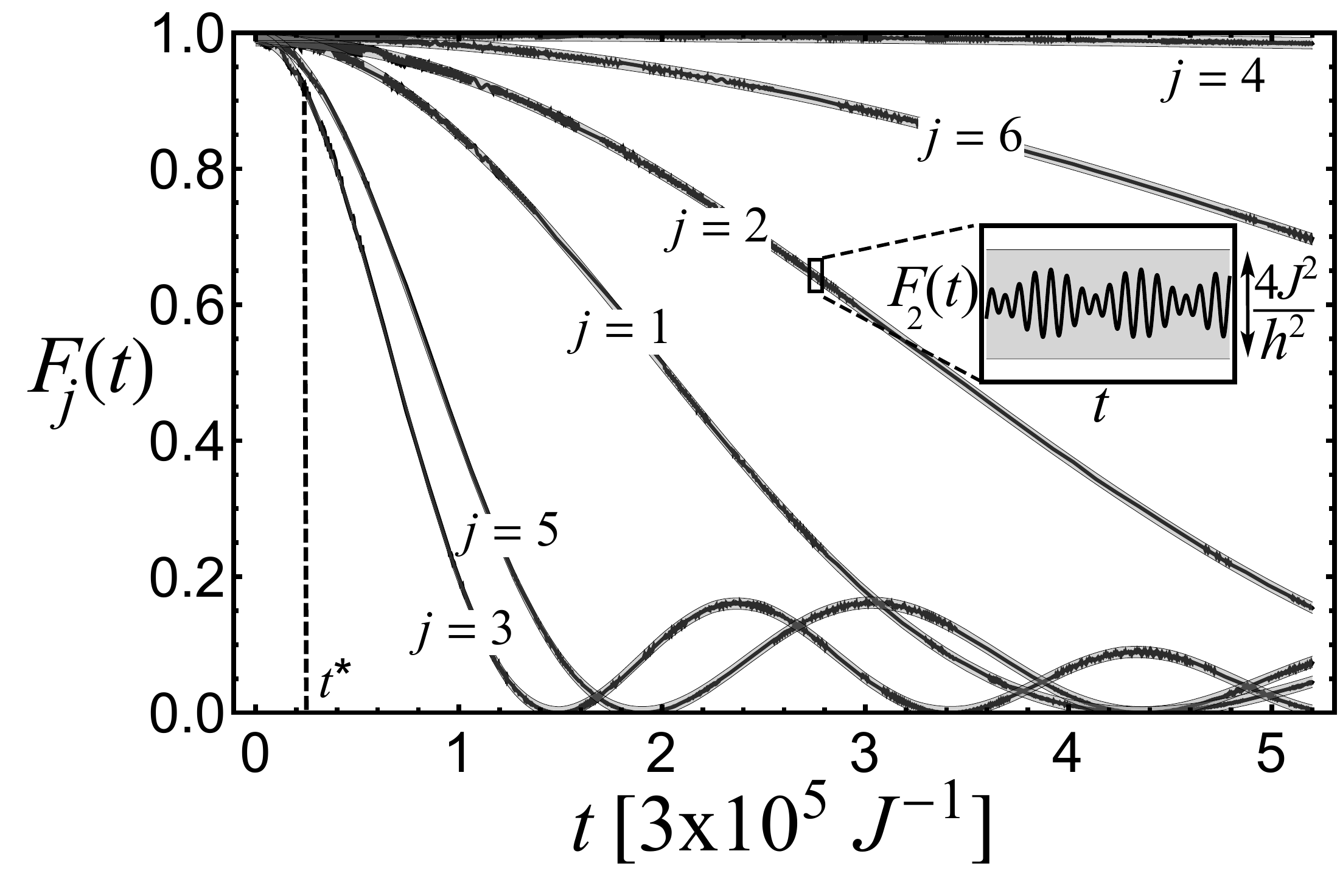}
\caption{Localization fidelity $F_j(t)$, Eq.~\eqref{eq:fidelity}, for a single particle initialized at site $j$ on a chain of length $L=15\times6=90$ with periodic boundary conditions; $h/J=20$. The width of the lines reflects the fast small-amplitude oscillations of $F_j(t)$ shown in the inset. The characteristic period of these oscillations is $4\pi/h$ for all values of $j=1,\ldots 6$.  The localization lifetime $t^*$ marked on the plot is given by Eq.~(\ref{eq:bound}), $t^{*}=1.3\times 10^{6}h^{-1}$ for the chosen $h/J$.
}
\label{fig:fidelity}
\end{figure}

The localization lifetime can be characterized by the time-dependent survival probability $F_j(t)$ on site $j$, i.e., the probability for particle to stay on the site where it was initially prepared,
\begin{equation}\label{eq:fidelity}
F_j(t)=\langle \hat n_j(t) \rangle_{n_j(0)=1}, \qquad \hat n_j=c_j^\dagger c_j.
\end{equation}
Function $F_j(t)$ has the meaning of localization fidelity, and we use this term in what follows.

In the tight-binding picture Eqs.~(\ref{eq:spec}) and (\ref{eq:energy_shift}) the time dependence of the site wave functions (Wannier wave functions) is well known. The localization fidelity is 
\begin{equation}
\label{eq:fidelity2}
F_j(t)=\left|(3/\pi)\int dk \,\exp[-i E_j(k) t]\right|^2=J_0^2\left(U_jt\right),
\end{equation}
where $J_0$ is the zero order Bessel function of the first kind. 

From Eq.~(\ref{eq:energy_shift}), function $F_j(t)$ scales with time as $(J/h)^5Jt$. The localization lifetime may then be defined as $t^* = {\rm const} J^{-1}(h/J)^5$. For concreteness we set, as in Eq.~\eqref{eq:bound},
%
$t^*= \tfrac{1}{50}\left(h/J\right)^5J^{-1}$.
%
Then the {\em minimal} over sites $j$ localization fidelity for $t<t^*$ is $F_{\rm min}(t^*)=\min_j F_j(t^*) \approx 0.93$. 

The localization fidelities for different sites are shown in Fig.~\ref{fig:fidelity}. They are obtained by diagonalizing the full Hamiltonian (\ref{eq:H}) in the absence of the inter-particle coupling. For the considered ratio $h/J=20$ the lines given by Eq.~(\ref{eq:fidelity2}) are within the linewidth of the numerical curves. This linewidth is $\sim J^2/h^2$ and is due to fast oscillations of the wave function between the neighboring sites with period $\sim h^{-1}$. On short time-scales $(t\ll t^*)$ these oscillations have the form  
\begin{equation}\label{eq:short-time}
F_j(t)\approx 1-\sum_{\alpha=\pm1}\frac{J^{2}}{(\varepsilon_{j}-\varepsilon_{j+\alpha})^{2}}\textrm{sin}^{2}\left(\frac{\varepsilon_{j}-\varepsilon_{j+\alpha}}{2}t\right).
\end{equation}
As seen from this expression, the localization fidelity remains close to unity at early times. Using Eq.~\eqref{eq:seq} we can obtain a lower bound for $F_{\min}(t)=\min_j F_j(t)$ at small times: $1-F_{\min}(t)<10 J^2/h^2$ for $t\ll t^*$. In the case $h/J=20$ shown in Fig.~\ref{fig:fidelity} this implies $F_{\min}(t)>0.975$ for such times [$F_{\min}(t)$ approaches the value 0.93 mentioned  above as $t$ approaches $t^*$].

These results show that, even in a periodic system, as a consequence of the small tunneling matrix elements $U_j$ between degenerate levels separated by 6 sites, Eq.~\eqref{eq:energy_shift}, the localization fidelity exhibits a large
separation of time and amplitude scales. In spite of the band structure of the overall spectrum, all particles remain localized to their initial sites for a parametrically large time. The narrow bandwidths $2|U_j|$ imply that the effective mass $m_*$  for translational motion is extremely large,  $m_*\propto h^5/J^6$. Not only does it make the particles very slow, but it also makes the long-time dynamics of the system highly susceptible to perturbations of the site energies. Random perturbations can lead to the Anderson localization of all eigenstates even where the typical magnitude of the fluctuations of $\ep_j$ is $W\gtrsim J^6/h^5$ for the considered period-6 chain. We note that for $W\lesssim J^6/h^5$ the localization lifetime $t^*$ is the same as in the absence of disorder. This is because the localization length (of stationary states) in this case extends over many unit cells.

For stronger disorder, $J^6/h^5\ll W\ll h$, the localization length of stationary states becomes shorter than period 6 and approaches inter-site distance with increasing $W$. However, for $W\sim h$, the disorder starts mixing localized states on neighboring sites. This leads to random local resonances and a sharp decrease of the localization lifetime.

\section{Many-body localization lifetime}
\label{sec:many-particle}

In this section we show that the localization lifetime $t^*$ scales as $J^{-6}$ even in the presence of many-particle transitions, if we use the energy sequence \eqref{eq:seq} and if the parameter of the interaction between the particles  is $\Delta\lesssim 1$. It is important that the single-site localization fidelity $F_j(t)$, Eq.~(\ref{eq:fidelity}), does not fall off with the increasing size of the system. 
We calculate $F_j(t)$ using a perturbative approach. It is based on finding the amplitudes of transitions between Fock states $|\{m_l\}\rangle$ that have equal energies and the same total number of particles ($m_l=0,1$ is the occupation of site $l$). Such resonant transitions are responsible for the decay of the initial configuration of occupied states, and thus of $F_j(t)$, because, as a result of a transition, the population changes by a factor $\sim 1/2$. On the contrary, in the absence of resonant transitions, site populations perform small-amplitude oscillations of the type of those described by Eq.~(\ref{eq:short-time}).

States $|\{m_l\}\rangle$ are eigenstates of the Hamiltonian $H_0$ defined in Eq.~(\ref{eq:H}). The transitions between the states are due to the operator $V$ in Eq.~(\ref{eq:H}). We treat this operator as a perturbation. 

In terms of the on-site Green function, $G_0(E)=(E-H_0)^{-1}$, the amplitude $T_{M;k}$ of a resonant transition between two many-body states $|\{m_l\}\rangle$ and $|\{m_l^\prime\}\rangle$ at order $k$ in $V\propto J$ that involves moving $M$ particles is
\begin{equation}
\label{eq:amplitude}
T_{M;k}(\{m_{l'}\} |\{m_l\})=\langle\{m_l^\prime\}|V(G_0V)^{k-1}|\{m_l\}\rangle,
\end{equation}
In this equation the energies of the states $|\{m_l\}\rangle$ and $|\{m_{l'}\}\rangle$ in the absence of hopping are assumed to be the same, 
$E=\langle\{m_l\}|H_0|\{m_l\}\rangle=\sum_l\left( m_l\varepsilon_l +  m_l m_{l+1}J\Delta\right) =\langle\{m_{l'}\}|H_0|\{m_{l'}\}\rangle$. 

Equation (\ref{eq:amplitude}) describes a process where the state $|\{m_l\}\rangle$, by being perturbed to the $(k-1)$th order of the perturbation theory in $V$, is brought in resonance with state $|\{m_{l'}\}\rangle$ and is separated from it by just one site. The operation of moving a particle over this site is performed by operator $V$. Clearly, Eq.~(\ref{eq:amplitude}) is symmetric with respect to which of the states $|\{m_l\}\rangle$ or $|\{m_{l'}\}\rangle$ is perturbed. The transition amplitude (\ref{eq:amplitude}) is of primary  importance for our analysis, as its maximal value determines the localization lifetime. 

As mentioned in Sec.~\ref{sec:seq}, to calculate the order of the transition amplitude $T_{M;k}$ in $J/h$ one has to include the number of particle-particle scattering where, in the course of virtual transitions, the particles find themselves on neighboring sites.  If the coupling $\Delta=0$, a many-particle transition would not occur even for $\sum_lm_l\ep_l = \sum_{l'}m_{l'}\ep_{l'}$ because of the destructive interference of the different sequences of virtual transitions. This is easy to see from Eq.~(\ref{eq:amplitude}), but is also clear from the qualitative argument: if the particles ``do not know'' of each other, they cannot make a resonant transition.

The leading-order term in the transition amplitude (\ref{eq:amplitude}) is thus constrained by the need to bring particles to the neighboring sites and allow them to scatter of each other. Sequences of virtual transitions where the number of such scattering is higher involve more transitions and thus give higher-order terms in the transition amplitude. Clearly, for an $M$-particle transition the minimal number of particle-particle scattering events is $M-1$. For $\Delta \lesssim 1$ they give a factor $\propto (J\Delta/h)^{M-1}$ in $T_{M;k}$, since the scattering amplitude is $\propto J\Delta$.
We show  below that, because of this factor, in our period-6 sequence, calculating $F_j(t)$ for  $t<t^*\propto J^{-1}(h/J)^5$ requires taking into account only transitions in which  one, two, or three particles are involved. Our analytical results are supported by numerical diagonalization of a half-filled 12-site chain described in Sec.~\ref{sec:numerical}.

\subsection{Two-particle resonances}
\label{sec:two-particle}

Here we consider resonant transitions between states differing by the occupation numbers at sites $(i,j)$ and $(i^\prime,j^\prime)$, that is, transitions that involve moving two particles from $(i,j)$ to $(i',j')$. The presence of other particles can shift the site energy levels by $J\Delta$. If this shift is small compared to the level spacing, i.e., for small $J\Delta/h$, it can be disregarded. The effect of this shift will be discussed later in Sec.~\ref{sec:hartree}.  

The feature of the energy sequence \eqref{eq:seq}, that underlies its design, is that it eliminates two-particle resonances separated by less than 6 steps (virtual transitions). A search of resonant two-particle energy combinations shows that the resonant transitions that involve the lowest number of steps are (a) $(6,12)\leftrightarrow(8,10)$ and (b) $(3,9)\leftrightarrow(5,7)$. Other leading-order two-particle transitions are related to these by a trivial shift by a multiple of 6 sites. The corresponding transition amplitudes are $T_{2;6}^{(a,b)}\propto  (J/h)^6 J\Delta $. Here, the factor $(J/h)^4$ comes from the minimal number of transitions to get from the initial to the final state. An extra factor $(J/h)^2$ comes from the two extra virtual transitions needed to bring the particles to neighboring sites and take them back to the resonating sites. The factor $J\Delta$ comes from the scattering amplitude. Numerically, from Eqs.~(\ref{eq:seq}) and (\ref{eq:amplitude})
\begin{equation}
\label{eq:2-amp2}
|T_{2;6}^{(a)}|=\frac{1152 }{385}\frac{J^6}{h^6}J\Delta ,\quad
|T_{2;6}^{(b)}|=\frac{10368}{385}\frac{J^6}{h^6}J\Delta.
\end{equation}

From Eq.~\eqref{eq:2-amp2} one sees that, for $J\Delta/h\lesssim 1$, the leading order two-particle transition amplitudes are parametrically smaller than the single-particle transition amplitudes, which are given by the bandwidths $U_j$ in Eq.~(\ref{eq:energy_shift}). As a result, the decay of site occupations $n_j(t)$  due to two-particle transitions is slower than due to single-particle transitions. As we shall see in Sec.~\ref{sec:hartree}, this remains true even for larger interactions $\Delta\sim h/J$. 

\subsection{Three-particle resonances}
\label{sec:three-particle}

We will now show that, for our periodic  sequence of site energies, resonant transitions that involve more than two particles also have small amplitudes, leading to a slow decay of the site populations $n_j(t)$. Indeed, for a 4-particle transition, the amplitude $T_{4;k}$ already has a factor $(J\Delta/h)^3$ from the scattering of particles off each other. In addition, one transition per the involved particle gives a factor $J^4$. Overall, we see that $T_{4;k}$ scales utmost as $J(J/h)^3(J\Delta/h)^3$. Clearly, for more than 4 particles the transition amplitudes will be of still higher order in $J/h$. Therefore, with account taken of the previous results, it is sufficient to study 3-particle transitions to make sure that the overall transition amplitude in a many-particle system scales as $J(J/h)^5$ for $\Delta \lesssim 1$.

From the above counting argument, the amplitude of a three-particle resonant transition could scale as $J(J/h)^2(J\Delta)^2$ if there were resonances between site energies that involve 3 transitions. However, Eq.~\eqref{eq:seq} specifically eliminates  processes at this order. Indeed, the maximal distance between the sites involved in a 3-particle transition with amplitude $\propto J^5$ is 6, it does not exceed the lattice period $N=6$, whereas the sequence (\ref{eq:seq}) has no three-particle resonances within the period. Therefore, to the leading order, the three-particle transition amplitude can be $\propto J(J/h)^3(J\Delta/h)^2$ or smaller. For $\Delta\lesssim 1$ this is the same scaling as for the amplitude of the single-particle transition. 

To determine whether resonant 3-particle transitions with amplitude $\propto J^6$ may take place for the sequence \eqref{eq:seq}, one has to analyze the three-particle energy combinations, $\varepsilon_i+\varepsilon_j+\varepsilon_k$. There occur two relevant transitions: (a) $(4,5,8,11)\leftrightarrow(4,6,7,9)$ and (b) $(1,4,7,8)\leftrightarrow(3,5,6,8)$. A static fourth particle on sites 4 and 8 in cases (a) and (b), respectively, is added to balance the interaction energy $J\Delta$ in the initial and final states. Equations~\eqref{eq:seq} and Eq.~\eqref{eq:amplitude} give, to the leading order in $J$,
\begin{equation}\label{eq:3-amp}
|T^{(a)}_{3;4}|=\frac{7776}{245} \frac{J^5}{h^5} J\Delta^2,\quad |T^{(b)}_{3;4}|=32 \frac{J^5}{h^5}J\Delta^2.
\end{equation}

From Eqs.~(\ref{eq:energy_shift}) and (\ref{eq:3-amp}), the localization lifetime, which is determined by the reciprocal maximal transition amplitude, scales as  $t^*\propto h^5/J^6$. This is consistent with the numerical data discussed in Sec.~\ref{sec:numerical} and presented in Fig.~\ref{fig:fidelity2}. The data shows no indications of a loss of localization fidelity due to processes that involve transitions of more than 3 particles over time $t^*$, even up to the regime of a comparatively strong interaction $\Delta\lesssim1$. Instead, we find that the limit on the localization lifetime with the increasing coupling parameter $\Delta$ is imposed by the coupling-induced quasi-single-particle resonances.

\subsection{Coupling-induced single-particle resonances}
\label{sec:hartree}

A sufficiently strong coupling can bring in resonance site energies for relatively close sites. This enables resonant single-particle transitions with comparatively large amplitude, as they involve a small number of virtual transitions. The energy shift of a site due to the presence of a particle on a neighboring site is $J\Delta$. Therefore, with increasing coupling parameter $\Delta$, there may emerge coupling-induced resonances of the site energies. For our sequence (\ref{eq:seq}) this happens first when the minimal single-particle energy difference, $h/12$, is matched by $nJ\Delta$, where $n=1,2$ is the number of neighboring particles around a given site. One can think of these transitions as 2- or 3-particle transitions, respectively, although only one particle moves between the sites. We therefore call them quasi-single-particle transitions. In contrast to the previous analysis, this is a non-perturbative situation: $J\Delta$ is not small compared to the single-particle level spacing. 

The transition amplitude is particularly large for the quasi-single-particle transitions (a) $(2,5)\leftrightarrow (2,3)$ and (b) $(6,9)\leftrightarrow (6,7)$; as before, the pair $(i,j)$ indicates the occupied sites in the periodically extended sequence (\ref{eq:seq}). For the considered here case $\Delta>0$ the particle that does not move is next to the site with a lower single-particle energy. To have the resonance for $\Delta<0$, the non-moving particle must be next to the site with higher energy. 

The strong-coupling transition amplitudes for resonances (a) and (b) are, respectively, 
\begin{equation}
\label{eq:amp1}
|T_{1;2}^{(a)}| = \frac{3}{10}\frac{J^2}{h},\quad |T_{1;2}^{(b)}|=\frac{3}{5}\frac{J^2}{h} \quad(J\Delta\approx h/12).
\end{equation}
Equation~\eqref{eq:amp1} shows that, for some particle configurations and for $J\Delta\approx h/12$, the localization lifetime sharply drops from the scaling (\ref{eq:bound}) to $t^* \sim h/J^2 \ll  h^5/J^6$. This resonant effect is seen in Fig.~\ref{fig:fidelity2} below. 

The transition amplitude for the quasi-single-particle resonance $\min|\ep_i-\ep_j| = 2J\Delta$ requires more virtual transitions than the resonance $\min|\ep_i-\ep_j| = J\Delta$ that leads to Eq.~(\ref{eq:amp1}). This is a consequence of the one-dimensional nature of the considered energy sequence. If one site has no neighbors while the other has two neighbors, the separation of the sites must be larger than two. For the sequence (\ref{eq:seq}), where $\min|\ep_i-\ep_j|=h/12$, the leading-order resonant transition amplitudes for $J\Delta \approx h/24$ are of fourth order in $J$. The relevant transitions are (a) $(3\leftrightarrow7)$, with the occupation of the surrounding sites $n_2=n_4 =0, n_6=n_8=1$,  and (b) $(5\leftrightarrow9)$, with the occupation of the surrounding sites $n_4=n_6=0, n_8=n_{10} = 1$.
Working out the transition amplitudes gives
\begin{equation}
\label{eq:1-amp}
|T_{1;4}^{(a)}|=\frac{672}{935}\frac{J^4}{h^3},\quad |T_{1;4}^{(b)}|=\frac{192}{91}\frac{J^4}{h^3} \quad(J\Delta\approx h/24).
\end{equation}
The localization lifetime $t^*$ for the resonance $J\Delta = h/24$, which is on the order of the inverse of the transition amplitude,  is longer than for $J\Delta = h/12$ but still much smaller than Eq.~(\ref{eq:bound}) in the absence of coupling-induced resonances. 

The transition amplitude determines the characteristic width of the range of $\Delta$ where the coupling-induced resonance is manifested. Indeed, the width of this resonance is determined by the condition that $J\Delta$ is on the order of the transition amplitude. This is in agreement with the data in Fig.~\ref{fig:fidelity2} discussed below. 

The analysis leading to Eqs.~(\ref{eq:amp1}) and \eqref{eq:1-amp}, combined with the results of Secs.~\ref{sec:single-particle} and \ref{sec:many-particle}, demonstrates that, for $h > 24 J$, all particles in the chain remain localized on their initially occupied sites up to time exceeding $t^*\sim 10^5 J^{-1}$. The estimate holds for any value of the coupling parameter $\Delta\leq1$. This is an important feature of the site energy sequence (\ref{eq:seq}). It was used in Eq.~\eqref{eq:bound}.

\subsection{Numerical results}
\label{sec:numerical}

Numerical studies of the localization lifetime in a many-particle system are invariably limited to a comparatively small number of particles. A convenient characteristic for such studies is function 
\begin{equation}
\label{eq:fid2}
\tilde F(t)=\left\langle\prod_{j\in \{j_0\}} \hat n_j(t)\right\rangle,
\end{equation}
where the set $\{ j_0\}$ enumerates the initially populated states, $n_j(0)=1$ for $j\in\{j_0\} $ and $n_j(0)=0$ otherwise. On the time scale where all populations $\langle n_j(t)\rangle_{n_j(0)=1}$ remain close to 1
\begin{align}
\label{eq:F_approximate}
\tilde F(t)\approx &1-\sum_{j\in \{j_0\}}\left[1- \langle n_j(t)\rangle_{n_j(0)=1}\right] \nonumber\\
&\equiv 1-\sum_{j\in \{j_0\}}\left[1-F_j(t)\right],
\end{align}
where $F_j(t)$ is the localization fidelity for site $j$ defined in Eq.~(\ref{eq:fidelity}). Equation (\ref{eq:F_approximate}) applies for  $1-F_j(t)\ll 1$ for $j\in \{j_0\}$.
\begin{figure}[h!]
\centering
\includegraphics[width=1.\columnwidth]{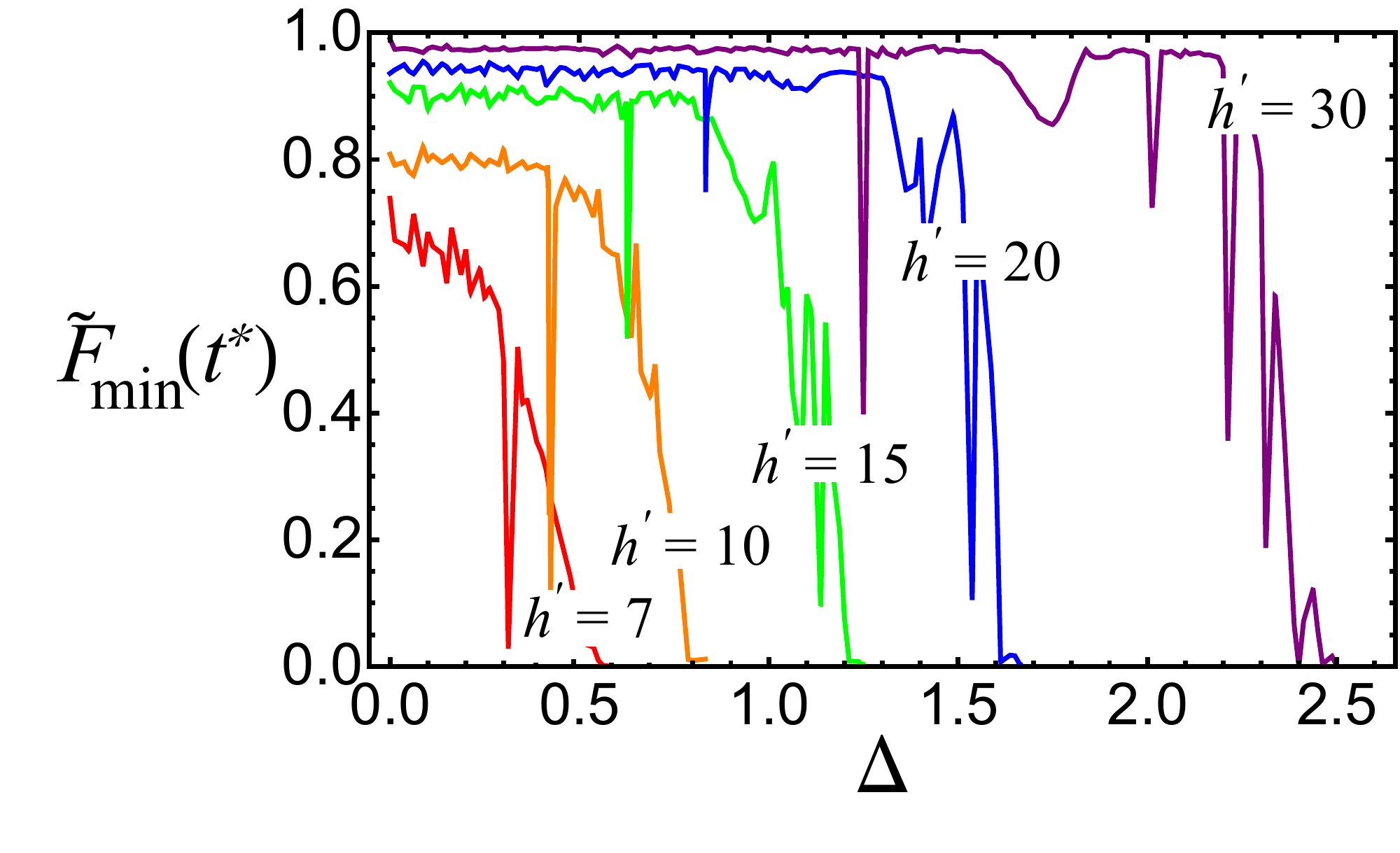}
\caption{The characteristic many-body localization fidelity as function of the particle-particle coupling parameter $\Delta$. Function  $\tilde F_{\min}(t^*)$ is given by the minimum of function $\tilde F(t^*)$,  Eq.~(\ref{eq:fid2}), with respect to all initial many-body states $|\{n_j\}\rangle$ for a half-filled 12-site chain. The characteristic localization lifetime $t^*$ for the considered system is defined in Eq.~\eqref{eq:bound}. Parameter $h'=h/J$ gives the ratio of the bandwidth of single-particle energies to the hopping amplitude. The results are invariant under the transformation $\Delta\to-\Delta$. We truncate each curve for $J\Delta>h/12$ for clarity. The strong suppression of $F_{\rm min}(t^*)$ near $\Delta\sim h/12,\,h/24$ results from coupling-induced quasi-single-particle resonances discussed in Sec.~\ref{sec:hartree}.}
\label{fig:fidelity2}
\end{figure}

From Eq.~(\ref{eq:F_approximate}), function $\tilde F(t)$ is an integrated characteristic of the loss of localization fidelity by all particles in the system. This makes it convenient for numerical studies of small systems. In contrast, for large systems $\tilde F(t)$ goes to zero even where all $F_j(t)$ are just weakly decreased from $F_j=1$ by the small nonresonant ``leakage'' to neighboring sites, cf. Eq.~(\ref{eq:short-time}). 

To reveal the effect of the particle-particle interaction on the localization lifetime, we present in Fig.~\ref{fig:fidelity2} the {\it minimal} value of function $\tilde F(t)$ for a half-filled 12-site chain described by Hamiltonian (\ref{eq:H}) and (\ref{eq:seq}). The results are obtained by diagonalizing the Hamiltonian numerically. The minimum is taken with respect to all initial distributions of the particles over the sites, $\tilde F_{\min}(t) = \min_{\{j_0\} } \tilde F(t)$, where $\{j_0\}$ gives the configuration of the six initially occupied sites. Therefore $\tilde F_{\min}(t)$ characterizes the worst case scenario in terms of the localization fidelity. Function $\tilde F_{\min}(t)$ is evaluated for the time $t=t^*\propto h^5/J^6$, where $t^*$ is  given by Eq.~\eqref{eq:bound}. Therefore $\tilde F_{\min}(t^*)$ is a direct test of the applicability of the theory.

Figure~\ref{fig:fidelity2} shows that $\tilde F_{\min}(t^*)$ is very close to 1 in a broad range of the coupling strength parameter $\Delta$ already for $h/J > 15$. This means that the many-body wave function at time $t<t^*$ remains close to the initial state, i.e. {\em all} particles remain near their initial sites. The small-$\Delta$ value of $\tilde F_{\min}$ is well described by the single-particle theory, cf. Fig.~\ref{fig:fidelity}, but in fact $\tilde F_{\min}(t^*)$ remains weakly dependent on $\Delta$ in a broad range of $\Delta$. This is to be expected, as for $J\ll h$ and $\Delta \lesssim 0.3$ the single-particle processes provide the strongest bound on the localization lifetime. This bound is only weakly affected by 3-particle transitions for somewhat larger $\Delta$, cf. Eqs.~(\ref{eq:energy_shift}) and (\ref{eq:3-amp}).

A dramatic drop of $\tilde F_{\min}(t^*)$ occurs for very strong coupling where $J\Delta$ approaches the minimum site energy difference $h/12$. As explained in Sec.~\ref{sec:hartree}, this drop is a consequence of the coupling-induced  resonances between site energies separated by two sites. One can also see the expected narrow dips of $\tilde F_{\min}$ for $J\Delta=h/24$. They correspond to the coupling-induced resonances between site energies separated by 4 sites and are therefore much weaker than the resonance for $J\Delta = h/12$. These resonances obviously involve ``fine tuning''. Overall the numerical results demonstrate robust long localization lifetime in the considered strongly coupled many-particle system.

\section{Conclusion}\label{sec:conclusion}

In this paper we have examined the dynamics of a one-dimensional many-body system from the perspective of the localization lifetime $t^*$, i.e., the time during which all particles (or equivalently, all excitations) remain on the sites they were initially placed, with a very large probability. To achieve a long localization lifetime one may not rely on the randomness of site energies, rather the ``disorder'' has to be constructed. We have proposed a simple sequence of site energies that, even for a strong particle-particle interaction,  makes the amplitudes of all resonant many-particle transitions scale as a high power of the ratio of the nearest-neighbor hopping amplitude $J$ to the single-particle energy bandwidth $h$. The localization lifetime is determined by the inverse of this amplitude. 

Somewhat unexpectedly, it turned out that the site energies can be chosen as a periodic sequence. For the considered model with nearest-neighbor hopping and short-range coupling, our period-6 sequence provides the same scaling for the many- and single-particle localization lifetime, $t^*\propto J^{-1}(h/J)^5$. The periodicity of the sequence has made it possible to study the localization lifetime analytically. The numerical results have confirmed the theory.

 A long localization lifetime and its scaling as a high power of $h/J$ is advantageous for applications of many-body systems in quantum information. It shows that, even where the coupling cannot be switched off, excitations can remain for a long time on the sites they were created and can be read out, while two-qubit operations can be done over a much shorter time $\sim J^{-1}$. In this context, a small delocalization $\sim (J/h)^2$ related to nonresonant inter-site hybridization of excitations should be compared to the error of the control and readout and can be reduced by changing $J/h$. 
 
Even though the present formulation differs from the conventional problem of many-body localization, the results are closely related to this problem. Indeed, since the amplitudes of all resonant many-particle transitions can be made very small even in a periodic system, already a weak site disorder, with width $\ll h$, will break the periodicity and could be sufficient for localization of the stationary states. The study of such localization is beyond the scope of this paper. Meanwhile, it is clear from our analysis that a weak site disorder will not affect the localization lifetime.

\begin{acknowledgements}

M. Schecter thanks the Villum Foundation for support. M. Shapiro and M. I. Dykman acknowledge partial support by the US National Science Foundation  through grants DMS-1362352 and DMR-1514591, respectively.

\end{acknowledgements}

\bibliographystyle{apsrev4-1}

%

\end{document}